\documentclass[journal]{IEEEtran}

\usepackage{amsmath}
\usepackage{amssymb}
\usepackage{algorithm}
\usepackage{algpseudocode}
\usepackage{graphicx} 
\usepackage{verbatim}

\begin{document}
\pagestyle{plain}

\title{CT Material Decomposition using \\Spectral Diffusion Posterior Sampling}
\author{Xiao~Jiang,
        Grace~J.~Gang,
        and~J.~Webster~Stayman \vspace{-.4in}
\thanks{Xiao~Jiang and J.~Webster~Stayman are with the Department of Biomedical Engineering, Johns Hopkins University, MD 21205 USA (e-mail: xjiang43@jhu.edu, web.stayman@jhu.edu).
Grace~J.~Gang is with the Department of Radiology, University of Pennsylvania. PA 19104 USA (e-mail: grace.j.gang@pennmedicine.upenn.edu).}
}

\maketitle

\vspace{-.5in}

\begin{abstract}
In this work, we introduce a new deep learning approach based on diffusion posterior sampling (DPS) to perform material decomposition from spectral CT measurements. This approach combines sophisticated prior knowledge from unsupervised training with a rigorous physical model of the measurements. A faster and more stable variant is proposed that uses a ``jumpstarted'' process to reduce the number of time steps required in the reverse process and a gradient approximation to reduce the computational cost. Performance is investigated for two spectral CT systems: dual-kVp and dual-layer detector CT. On both systems, DPS achieves high Structure Similarity Index Metric Measure(SSIM) with only $10\%$ of iterations as used in the model-based material decomposition(MBMD). Jumpstarted DPS (JSDPS) further reduces computational time by over $85\%$ and achieves the highest accuracy, the lowest uncertainty, and the lowest computational costs compared to classic DPS and MBMD. The results demonstrate the potential of JSDPS for providing relatively fast and accurate material decomposition based on spectral CT data.
\end{abstract}


\section{Introduction}

\IEEEPARstart{S}{pectral} CT has enabled a number of applications including density estimation, virtual monochromatic imaging, and contrast agent enhancement.
Many of these applications rely on material decomposition, which reconstructs basis material density maps from the spectral projections\cite{long2014multi}. Material decomposition is an ill-conditioned nonlinear inverse problem without an explicit solution. Existing material decomposition algorithms can be categorized into three main types: analytical decomposition\cite{jiang2021fast}, iterative/model-based decomposition\cite{tilley2019model}, and learning-based decomposition\cite{gong2020deep}. 

Deep learning methods have been extensively used in medical image formation including spectral CT\cite{gong2020deep,wu2021deep,abascal2021material,zhu2022feasibility}. Such approaches leverage prior knowledge learned from large datasets and can generally surpass classic approaches in terms of image quality. However, many deep learning methods do not directly leverage a physical model of the measurements, raising concerns about robustness, network hallucinations, etc. Some methods integrate a physical model as part of network training to improve the data consistency\cite{fang2021iterative,eguizabal2022deep}. However, these trained networks are tailored to specific system models, requiring network retraining for system changes in protocol, technique, device, etc.

Diffusion Posterior Sampling (DPS)\cite{chung2022diffusion} has established a new framework to integrate a learned prior and physics model. Specifically, DPS starts with unsupervised training of a score-based generative model (SGM) to capture the target domain distribution, then application of a reverse process to estimate image parameters - alternating between SGM reverse sampling to drive the image towards the target distribution and model-based updates to improve the data consistency with measurements. As a result, the final output adheres to both the prior distribution and the actual measurements. A major advantage of DPS is that its network training is not specific to any physical model, allowing for application across different imaging systems. Recently, we have applied DPS to nonlinear CT reconstruction\cite{li2023diffusion}, demonstrating its effectiveness for both low-mA and sparse-view single-energy CT reconstruction. 

In this work, the original DPS framework is expanded to Spectral DPS (SDPS) specifically for material decomposition in spectral CT. To further enhance this method, we propose a ``jumpstarted'' sampling strategy with gradient approximation, which significantly stabilizes the sampling process and reduces computational costs. The performance of both SDPS and its jumpstarted variant (JSDPS) are demonstrated in dual-kVp\cite{cassetta2020fast} and dual-layer CT\cite{wang2021high}. 

\section{Methodology}
\subsection{Spectral Deep Posterior Sampling}
\label{sec:dps}
\subsubsection{Score-based Generative Model (SGM) for Inverse Problems}
A SGM\cite{song2020score} defines a forward process which continuously perturbs the target-domain sample with time-dependent noise. The corresponding reverse process generates new samples by inverting the perturbation process. Both forward and reverse processes are described by a stochastic differential equation (SDE). Specifically, the SDEs for a Denoising Diffusion Probabilistic Model (DDPM)\cite{ho2020denoising} have the following form at time $t$:
\begin{subequations}
\begin{equation}
\label{eq:forward}
    Forward: \mathrm{d}\textbf{x} = -\frac{\beta_t}{2} \textbf{x}\mathrm{d}t + \sqrt{\beta_t}\mathrm{d}\textbf{w}  
\end{equation}
\begin{equation}
\label{eq:reverse}
    Reverse: \mathrm{d}\textbf{x} = [-\frac{\beta_t}{2} \textbf{x}-\beta_t\nabla_{\textbf{x}_t} \mathrm{log} p_t({\textbf{x}_t})]\mathrm{d}t + \sqrt{\beta_t}\mathrm{d}\textbf{w}
\end{equation}
\end{subequations}
\noindent where $\beta_t$ is the time-dependent noise variance and $\mathrm{d}\textbf{w}$ is the standard Wiener process. The unknown score function $\nabla_{\textbf{x}_t} \mathrm{log}p_t({\textbf{x}_t})$ is approximated by a deep neural network $\textbf{s}_\theta(\textbf{x}_t,t)$. 
A SGM is also capable of posterior sampling from conditional distribution $p(\textbf{x}|\textbf{y})$:
\begin{equation}
\label{eq:condition}
    \mathrm{d}\textbf{x} = [-\frac{\beta_t}{2} \textbf{x}-\beta_t\nabla_{\textbf{x}_t} \mathrm{log} p_t({\textbf{x}_t|\textbf{y}})]\mathrm{d}t + \sqrt{\beta_t}\mathrm{d}\textbf{w}
\end{equation}
In the context of CT, $\textbf{y}$ and $\textbf{x}$ are the projection measurements and the image volume, respectively. Leveraging Bayes rule $p(\textbf{x}|\textbf{y}) \propto p(\textbf{x})p(\textbf{y}|\textbf{x})$, the reverse sampling \eqref{eq:condition} may be reformulated as:
\begin{equation}
\label{eq:bayes}
\begin{aligned}
    \mathrm{d}\textbf{x} = &[-\frac{\beta_t}{2} \textbf{x}-\beta_t\nabla_{\textbf{x}_t} \mathrm{log} p_t({\textbf{x}_t})]\mathrm{d}t + \sqrt{\beta_t}\mathrm{d}\textbf{w} \\
    &- \beta_t\nabla_{\textbf{x}_t} \mathrm{log} p_t({\textbf{y}|\textbf{x}_t})\mathrm{d}t
\end{aligned}\
\end{equation}
\noindent The conditional distribution $p_t({\textbf{y}|\textbf{x}_t})$ can be approximated\cite{chung2022diffusion} as
\begin{equation}\
\label{eq:dps_appro}
    p_t(\textbf{y}|\textbf{x}_t) \approx p_t(\textbf{y}|\hat{\textbf{x}}_0), \text{where} \ \hat{\textbf{x}}_0 = 
\frac{1}{\sqrt{\overline{\alpha}_t}}(\textbf{x}_t+(1-\overline{\alpha}_t))\textbf{s}_\theta(\textbf{x}_t,t)
\end{equation}

\noindent Based on this approximation, the deep posterior sampling process could be expressed as\cite{li2023diffusion}:\
\begin{equation}
\label{eq:dps_chain}
\begin{aligned}
    \mathrm{d}\textbf{x} = &[-\frac{\beta_t}{2} \textbf{x}-\beta_t\nabla_{\textbf{x}_t} \mathrm{log} p_t({\textbf{x}_t})]\mathrm{d}t + \sqrt{\beta_t}\mathrm{d}\textbf{w} \\
    &- \beta_t\nabla_{\hat{\textbf{x}}_0} \mathrm{log} p_t({\textbf{y}|\hat{\textbf{x}}_0}) \nabla_{\textbf{x}_t} {\hat{\textbf{x}}_0}\mathrm{d}t
\end{aligned}
\end{equation}

\noindent Note that the first two term is exactly same as the unconditional DDPM sampling \eqref{eq:reverse}, and $p_t(\textbf{y}|\hat{\textbf{x}}_0)$ is the likelihood term for the measurements. Therefore, the reverse sampling \eqref{eq:dps_chain} integrates the prior information captured by DDPM and the physical measurement model provided by the likelihood function. 

The forward process is discretized into $T$ time steps \cite{ho2020denoising}:
\begin{equation}
\label{eq:ddpm_f}
\begin{aligned}
    &\textbf{x}_t = \sqrt{\bar{\alpha}_t}\textbf{x}_0 + \sqrt{1-\bar{\alpha}_t}\boldsymbol{\epsilon} \\
    \text{where} &\ \bar{\alpha}_t=\prod_{i=1}^{t} (1-\beta_t), \ \boldsymbol{\epsilon} \sim \mathcal{N}(0,\boldsymbol{I}), \ t=1,2...T
\end{aligned}
\end{equation}

\noindent Since $\boldsymbol{\epsilon} = -\sqrt{1-\bar{\alpha}_t} \ \mathrm{log} p_t({\textbf{x}_t})$, DDPM trains a network $\boldsymbol{\epsilon}_\theta(\textbf{x},t)$ to predict the noise $\boldsymbol{\epsilon}$. That is, network parameters, $\theta$, are estimated

\begin{equation}
\label{eq:ddpm_train}
    \theta^* =  \text{argmin} \ \mathbb{E}_{\textbf{x}_0}\mathbb{E}_{\boldsymbol{\epsilon},t} \|\boldsymbol{\epsilon}_\theta(\textbf{x}_t,t)-\boldsymbol{\epsilon}\|_2^2
\end{equation}
\noindent Using the trained network and physical model, the image $\textbf{x}_0$ may be reconstructed by solving the reverse SDE\cite{ho2020denoising,song2020denoising}.

\subsubsection{Diffusion Posterior Sampling for Material Decomposition}
\begin{figure}[t]
\centering
\includegraphics[width=\linewidth]{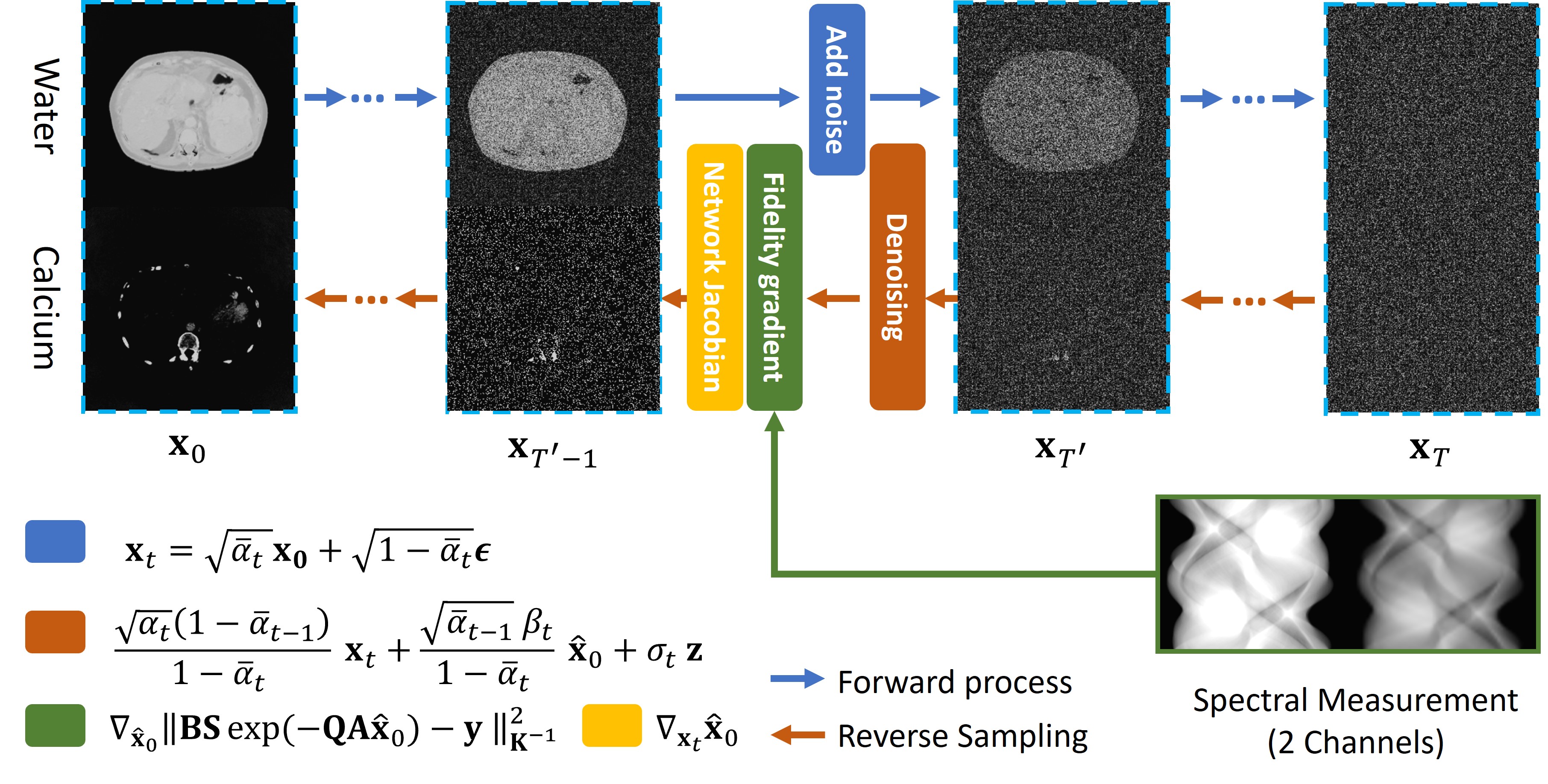}
\caption{SDPS Workflow: Basis material images are concatenated to form $\textbf{x}_0$. The forward process progressively perturbs images with noise, and a noise prediction network is trained. The reverse process uses DDPM sampling to progressively denoise the image and uses a likelihood-based update to enforce data consistency.}
\label{fig:dps} 
\end{figure}

This work aims to apply the DPS for the material decomposition problem, which estimates basis material densities from the spectral CT measurements\cite{long2014multi}. We adopt a general spectral CT model\cite{wang2021high} where measurements $\textbf{y}$ are assumed to follow a multi-variate Gaussian distribution with covariance $\textbf{K}$: $\textbf{y} \sim \mathcal{N}(\overline{\textbf{y}}, \textbf{K})$ and mean:
\begin{equation}
\label{eq:phyics_model}
    \overline{\textbf{y}} = \textbf{BS}\exp({-\textbf{QAx}})
\end{equation}
\noindent where $\textbf{x}$ are densities for each basis volume, $\textbf{Q}$ are mass attenuation coefficients for each basis, $\textbf{A}$ represents (channel-specific) projection matrices, $\textbf{S}$ are channel-specific spectral response, and $\textbf{B}$ captures overall gain effects. 
\noindent Substituting Eq.\eqref{eq:phyics_model} into Eq.\eqref{eq:dps_chain}, we have the spectral DPS (SDPS) for material decomposition:
\begin{equation}
\begin{aligned}
\label{eq:spectral_dps}
    \mathrm{d}\textbf{x} = &[-\frac{\beta_t}{2} \textbf{x}-\beta_t\nabla_{\textbf{x}_t} \mathrm{log} p_t({\textbf{x}_t})]\mathrm{d}t + \sqrt{\beta_t}\mathrm{d}\textbf{w} \\
    &- \beta_t\nabla_{\hat{\textbf{x}}_0} \|\textbf{BS}\exp({-\textbf{QA}\hat{\textbf{x}}_0})-\textbf{y}\|_{\textbf{K}^{-1}}^2 \nabla_{\textbf{x}_t} {\hat{\textbf{x}}_0} \mathrm{d}t
\end{aligned}
\end{equation}
The SDPS pseudo-code is shown below and illustrated in Fig. \ref{fig:dps}. 
\begin{algorithm}
\label{algo1}
\caption{Spectral Diffusion Posterior Sampling (SPDS)}
\begin{algorithmic}[1]\small
\State $T$: diffusion steps
\State $\eta_t$: step size
\State $\textbf{y}$: spectral CT measurement
\State $\textbf{x}_T\sim\mathcal{N}~(0,\boldsymbol{I})$
\For{\texttt{$t = T$ to $1$}}:
    \State $\textbf{z} \sim \mathcal{N}~(0,\boldsymbol{I})$
    
    \State $\hat{\textbf{x}}_0 = \frac{1}{\sqrt{\bar{\alpha}_t}}(\textbf{x}_t-\sqrt{1-\bar{\alpha}_t}\boldsymbol{\epsilon}_\theta(\textbf{x}_t,t))$
    
    \State $\textbf{x}_{t-1}' = \frac{\sqrt{\alpha_t}(1 - \bar{\alpha}_{t-1})}{1 - \bar{\alpha}_t} \textbf{x}_t + 
    \frac{\sqrt{\bar{\alpha}_{t-1}}\beta_t} {1 - \bar{\alpha}_t} \hat{\textbf{x}}_0 + \sigma_t \textbf{z}$
    
    \State $\textbf{x}_{t-1} = \textbf{x}_{t-1}' - \eta_t \nabla_{\textbf{x}_t} \left\| \textbf{BS}\exp({-\textbf{QA}\hat{\textbf{x}}_0}) - \textbf{y}\right\|_{\textbf{K}^{-1}}^2\nabla_{\textbf{x}_t} {\hat{\textbf{x}}_0}$
\EndFor
\end{algorithmic}
\end{algorithm}

\subsection{Strategies for Fast and Stable SPDS}
\label{sec:idps}
\subsubsection{Instability of SDPS}
The SDPS framework has the advantage of combining a learned prior and a physics model. However, we have observed that the stability of SDPS is highly dependent on the step size, $\eta_t$, and that even with careful tuning, decompositions present large variations over posterior samples and hallucinations. Careful design of a step size scheduler is required to balance the prior information and measurements, particularly for the earlier steps in the reverse process. In early steps, $\textbf{x}_t$ is far from the solution. While large step sizes may be favorable to enhance data consistency, this risks disrupting diffusion updates - making it challenging to pick optimal step sizes.

\subsubsection{Jumpstarted Sampling} 
For many imaging problems, an initial estimate is readily available through fast computation. For example, in spectral CT, projection-domain or image-domain material decomposition \cite{stenner2007empirical,mendoncca2013flexible} can be used as the first-pass estimate, which we denote as $\hat{\textbf{x}}_0^f$. Forward diffusion can be directly performed on $\hat{\textbf{x}}_0^f$ as $\hat{\textbf{x}}_t^f = \sqrt{\bar{\alpha}}_t\hat{\textbf{x}}_0^f  + (\sqrt{1-\bar{\alpha}_t})\boldsymbol{\epsilon}$. The difference between the distribution of $\hat{\textbf{x}}_t$ and $\hat{\textbf{x}}_t^f$ can be quantified by the KL divergence:
\begin{equation}
\label{eq:KL}
D_{KL}(p(\textbf{x}_t|\textbf{x}_0)\parallel p(\hat{\textbf{x}}_t^f|\hat{\textbf{x}}_0^f))=\frac{\bar{\alpha}_t}{2(1-\bar{\alpha}_t)}\|\hat{\textbf{x}}_0^f-\textbf{x}_0\|_2^2.
\end{equation}
\noindent In forward diffusion, ${\bar{\alpha}_t}/{(1-\bar{\alpha}_t)}$ progressively diminishes, ultimately converging towards zero. Consequently, Eq.\eqref{eq:KL} implies that we can expect for sufficiently large $t$, the difference between the distribution of $\textbf{x}_t$ and $\hat{\textbf{x}}_t^f$ is small enough to be ignored, then the reverse sampling can start from $\hat{\textbf{x}}_0^f$ instead of pure noise. In effect, we can skip early time steps by using an approximate solution to which an appropriate amount of noise has been added - yielding a more stabilized decomposition by using an initialization closer the solution as well as a faster solution having ``jumpstarted" over many early time steps. 

\subsubsection{Gradient Approximation}
SDPS computes the data fidelity gradient term via chain rule: $\nabla_{\textbf{x}_t} \mathrm{log} p_t({\textbf{y}|\hat{\textbf{x}}_0}) = \nabla_{\hat{\textbf{x}}_0} \mathrm{log} p_t({\textbf{y}|\hat{\textbf{x}}_0}) \nabla_{\textbf{x}_t}{\hat{\textbf{x}}_0}$. The $\nabla_{\textbf{x}_t}{\hat{\textbf{x}}_0}$ term may be expanded as:\vspace{-.1in}
\begin{equation}
\label{jacob}
\nabla_{\textbf{x}_t}{\hat{\textbf{x}}_0} = \frac{1}{\sqrt{\bar{\alpha}_t}}(1-\sqrt{1-\alpha_t}\nabla_{\textbf{x}_t}\boldsymbol{\epsilon}_\theta(\textbf{x}_t,t))\vspace{-.1in}
\end{equation}
\noindent The computation of the Jacobian $\nabla_{\textbf{x}_t}\boldsymbol{\epsilon}_\theta(\textbf{x}_t,t)$ is both time- and memory-intensive, particularly for high dimensional images and deep neural networks. We have experimentally observed that the Jacobian is well-approximated by a diagonal matrix for each time step, and the diagonal elements are of limited range. Therefore, we approximate the Jacobian by a constant (which could be further absorbed into the step size). Incorporating both the jumpstarted sampling and the gradient approximation,  we propose the jumpstarted spectral DPS (JSDPS), which is outlined in Algorithm 2 below.

\begin{algorithm}
\caption{Jumpstarted Spectral DPS (JSDPS)}
\begin{algorithmic}[1]\small
\State $T, T'$: training steps, sampling steps, $T'\ll T$
\State $\eta_t$: step size
\State $\textbf{y}$: spectral CT measurement
\State $\hat{\textbf{x}}_0^f$: image-domain decomposition

\State $\boldsymbol{\epsilon}\sim\mathcal{N}~(0,I)$
\State $\textbf{x}_{T'} = \hat{\textbf{x}}_0^{T'}=\sqrt{\bar{\alpha}_{T'}}\textbf{x}_0 + \sqrt{1-\bar{\alpha}_{T'}}\boldsymbol{\epsilon}$
\For{\texttt{$t = T'$ to $1$}}:
    \State $\textbf{z} \sim \mathcal{N}~(0,\boldsymbol{I})$
    
    \State $\hat{\textbf{x}}_0 = \frac{1}{\sqrt{\bar{\alpha}_t}}(\textbf{x}_t-\sqrt{1-\bar{\alpha}_t}\boldsymbol{\epsilon}_\theta(\textbf{x}_t,t))$
    
    \State $\textbf{x}_{t-1}' = \frac{\sqrt{\alpha_t}(1 - \bar{\alpha}_{t-1})}{1 - \bar{\alpha}_t} \textbf{x}_t + 
    \frac{\sqrt{\bar{\alpha}_{t-1}}\beta_t} {1 - \bar{\alpha}_t} \hat{\textbf{x}}_0 + \sigma_t \textbf{z}$
    
    \State $\textbf{x}_{t-1} = \textbf{x}_{t-1}' - \eta_t \nabla_{\hat{\textbf{x}}_0} \left\| \textbf{BS}\exp({-\textbf{QA}\hat{\textbf{x}}_0}) - \textbf{y}\right\|_{\textbf{K}^{-1}}^2$
\EndFor
\end{algorithmic}
\end{algorithm}

\subsection{Unconditional Multi-Material Generation}
\label{sec:ddpm}
\subsubsection{Dataset Generation}
This work focuses on two material decomposition with water and calcium as bases. To build spectral CT and material density dataset, we collected 32000 clinical chest CT slices from the CT Lymph Nodes dataset\cite{roth2014new}. Those images are pre-processed to convert Hounsfield Units to attenuation coefficients and remove patient beds. An upper bound value (2000HU) was also applied to remove large attenuation values from metal. Two soft threshold functions were applied to approximate water and calcium densities (unit: $g/ml$):
\begin{subequations}
\label{eq:material}
\begin{equation}  
 \rho_{w}(\mu) = 
\begin{cases}
    k_w\mu, & \text{if } \mu \leq \mu_w \\
    k_w\mu_w - k_{wc}(\mu-\mu_w), & \text{if } \mu_w < \mu < \mu_c\\
    0, & otherwise
\end{cases} 
\end{equation}\
\begin{equation}  
 \rho_{c}(\mu) = 
\begin{cases}
    k_c(\mu-\mu_c)+k_{cw}(\mu_c-\mu_w), & \text{if } \mu \geq \mu_c \\
    k_{cw}(\mu-\mu_w), & \text{if } \mu_w < \mu < \mu_c\\
    0, & otherwise\
\end{cases}    
\end{equation}
\end{subequations}

\noindent In this work, $k_w$, $k_{wc}$, $k_{cw}$, $k_{c}$ were empirically set to $5.18$, $-8.77$, $5.69$, $2.12~g/cm^2$, while $\mu_w$ and $\mu_c$ were $0.22~cm^{-1}$ and $0.35~cm^{-1}$. The water and calcium densities were used as ground truth to simulate spectral measurements. 

\subsubsection{Network Training}
DDPM\cite{ho2020denoising} is employed as the SGM framework, with the Residual Unet as the backbone network. Paired water and calcium images are concatenated to form the 2 channel image $\textbf{x}_0$. The discretized diffusion process uses $T = 1000$ time steps with a linear variance scheduler from $\beta_1 = 1e^{-4}$ to $\beta_{1000} = 0.02$. Training and validation datasets contain 25000 slices and 5000 slices, respectively. The loss function \eqref{eq:ddpm_train} is minimized by the Adam optimizer with a batch size of $16$ and a learning rate of $10^{-4}$. Training terminated after 200 epochs.

\subsection{Conditional Posterior Sampling for Different Spectral Systems}
\label{sec:appli}
The unconditional DDPM training captures the material bases distribution without specification of the spectral CT device. The trained model can be used in material decomposition for arbitrary spectral CT systems. Here we investigate the SDPS/JSDPS performance on a simulated dual-layer CT \cite{wang2021high} and dual-kVp CT system \cite{cassetta2020fast}. The dual-layer CT uses a detector with $300\mu m$ CsI and $600\mu m$ CsI scintillators in the top and bottom layers, respectively (with a 5mm gap between layers). The x-ray tube operates at 120~kVp. The dual-kVp system uses single-layer detector with $600\mu m$ CsI. The tube voltage alternates between 80~kVp and 120~kVp every other view. For both systems, 800 projections were simulated with Poisson noise equivalent to an exposure of 0.05 mAs/view. Voxel size and detector pixel size were set to $0.8~mm$ and $1.0~mm$, respectively.

\subsection{Evaluation}
We first evaluated the performance of JSDPS with and without gradient approximation. A standard image-domain decomposition from filtered backprojection reconstructions, i.e., the initialization of the JSDPS algorithm, was also shown for comparison. 
Second, we compared the performance of several material decomposition algorithms including model-based material decomposition (MBMD) \cite{tilley2019model}, SDPS, and JSDPS. MBMD was formulated as the following optimization problem:
\begin{equation}  
\label{eq:mbmd}
\hat{\textbf{x}} = \text{argmin} \|\textbf{BS}\exp({-\textbf{QAx}})-\textbf{y}\|_{\textbf{K}^{-1}}^2+\lambda_w\textbf{R}(\textbf{x}_w)+\lambda_c\textbf{R}(\textbf{x}_c)
\end{equation}
\noindent where $\textbf{R}$ is a quadratic gradient roughness penalty. Regularization strengths, $\lambda_w, \lambda_c$, for water and calcium were set to $10^{-4}$ and $4\times10^{-4}$, respectively. We used 10000 iterations of separable paraboloidal surrogates updates, and final change in loss is $\sim 0.01\%$ from the previous iteration. SDPS was implemented according to Algorithm 1, and the step size scheduler follows \cite{chung2022diffusion}: $\eta_t = \eta/\|\textbf{BS}\exp({-\textbf{QAx}})-\textbf{y}\|_{\textbf{K}^{-1}}$. We swept $\eta$ from $0.1$ to $10$, and the optimal $\eta$ was determined as the one with minimal MSE from ground truth. JSDPS used a constant step scheduler with $T'=150,\eta_t = 0.02$. Performance was evaluated for computation time (calculated for eight parallel decompositions), and image quality - quantified by Structural Similarity Index Measure(SSIM) and Peak Signal-to-Noise Ratio (PSNR).

Additionally, to compare variability and biases in the posterior samples provided by SDPS and JSDPS, we compute sample bias and standard deviation over an ensemble of 16 outputs:
\begin{equation}  
\begin{matrix}    
    \text{Bias} = \|\mathbb{E}\{\hat{\textbf{x}}\}-\textbf{x}\|_2 &
    \text{Std} = \|\mathbb{E}\{(\hat{\textbf{x}}-\mathbb{E}\{\hat{\textbf{x}}\})^2\}\|_2
    \end{matrix}
\label{eq:biasvar}
\end{equation}

\section{Results}
\label{sec:results}
\begin{figure*}[ht]
\centering
\includegraphics[width=\textwidth]{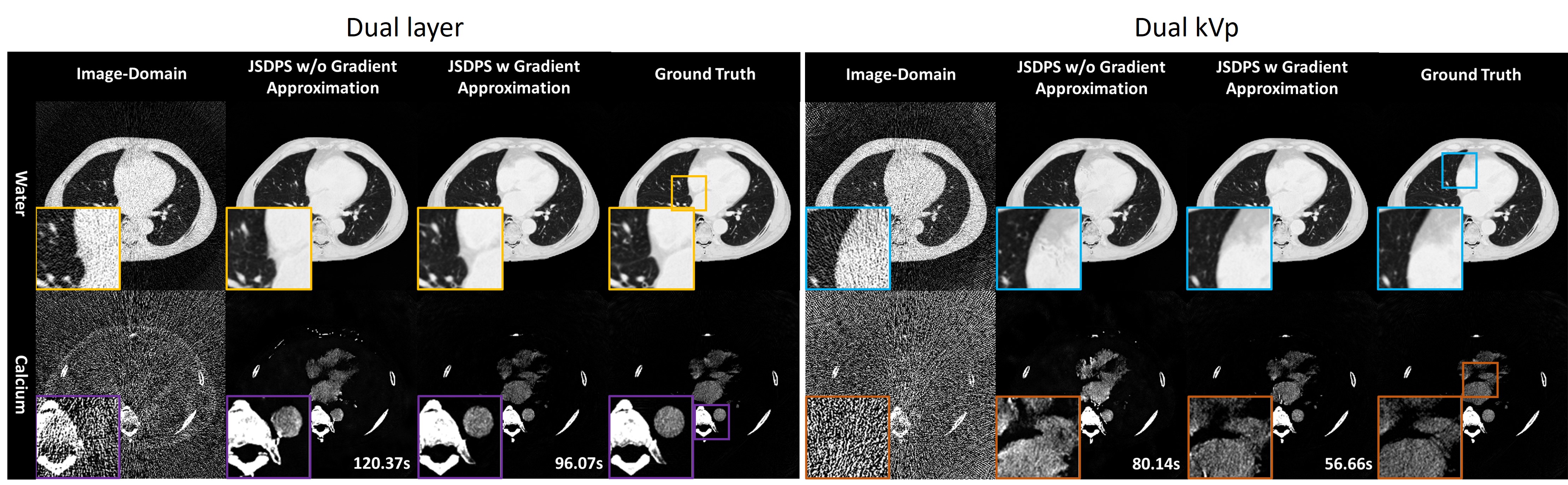}

\caption{Decomposed water (top, W/L: $1.2/0.6~g/ml$) and Calcium (bottom, W/L: $0.05/0.1~g/ml$) images. Left to right: Image-domain decomposition, JSDPS without gradient approximation, JSDPS with gradient approximation, ground truth. Computational time for eight outputs is provided in the bottom left corner.}
\label{fig:res1} 
\end{figure*}

\begin{figure*}[ht]
\centering
\includegraphics[width=\textwidth]{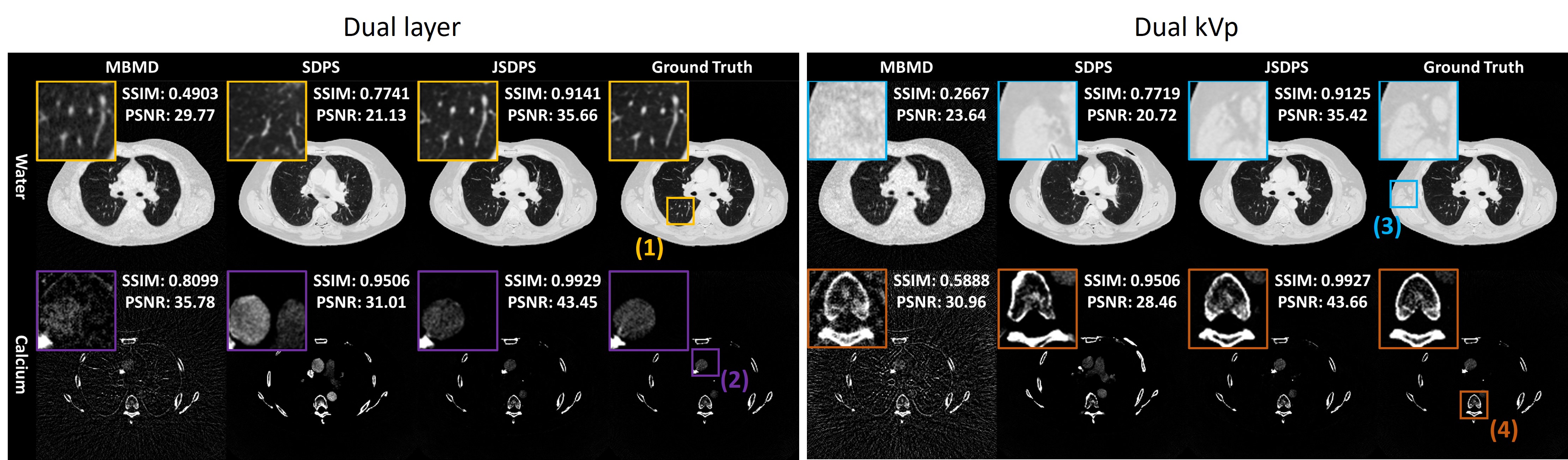}

\caption{Decomposed water (top, W/L: $1.2/0.6~g/ml$) and Calcium (bottom, W/L: $0.05/0.1~g/ml$) for (left to right): MBMD, SDPS, JSDPS, ground truth.}
\label{fig:res_compare} 
\end{figure*}

\subsection{Illustration of JSDPS variants}
Fig.\ref{fig:res1} displays JSDPS results as compared with standard image-domain decomposition (initialization) and the ground truth. Image-domain decomposition shows significant image noise and streaking. JSDPS without gradient approximation effectively captures most of the high-contrast structures, including smaller pulmonary vessels. However, as seen in the zoomed-in ROIs, low-contrast soft tissue hallucinations are present. In the contrast enhanced areas of the calcium image, such as the aorta, deformations are also observed. By using the gradient approximation, the reverse sampling achieves $20.2\%$ and $29.3\%$ computation time decrease for the dual-layer and dual-kVp system, respectively. Gradient approximation also effectively mitigates the hallucinations and enhances the consistency with ground truth for vascular structures.

\begin{figure}[ht]
\centering
\includegraphics[width=\columnwidth]{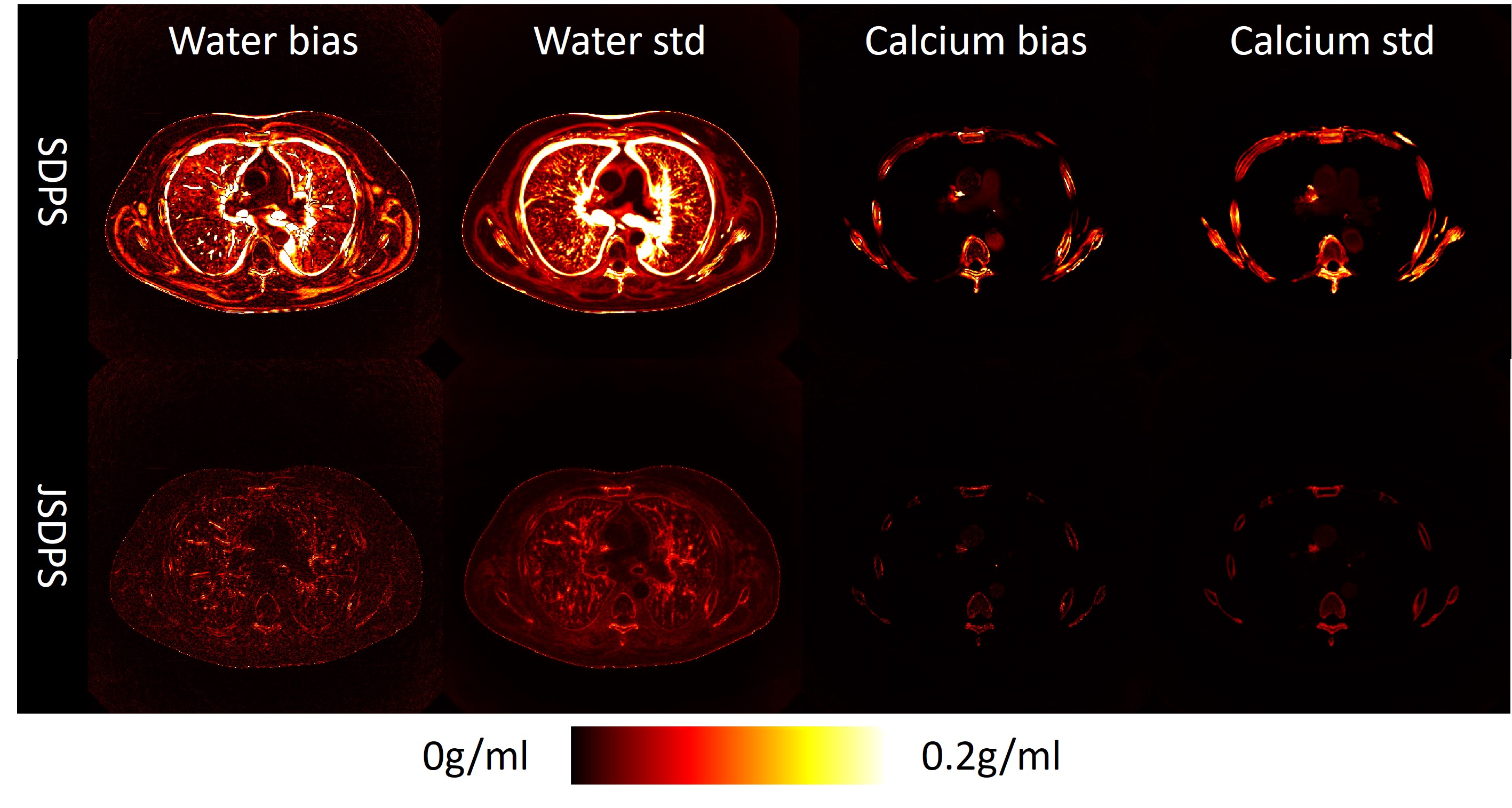}
\caption{Bias and standard deviation map for SDPS and JSDPS. Here we only show the results of dual-kVp system, which is similar to that of the dual-layer system.}
\label{fig:res_std} 
\end{figure}

\subsection{Comparison of SDPS, JSDPS, and MBMD}
\subsubsection{Image Quality}
Figure \ref{fig:res_compare} compares the performance of SDPS, JSDPS, and MBMD in water and calcium bases for the dual layer and dual kV systems. Four zoomed in areas are shown in the (1) lung, (2) heart (iodine-filled aorta in the calcium basis), (3) low-contrast soft tissue, and (4) spine. Comparing across all three algorithms, MBMD produces images with the highest noise (evident in ROIs 2 and 3) at the lowest spatial resolution (evident in ROI 1). The SDPS results have low noise and high spatial resolution, but are prone to hallucinations observed as erroneous anatomy in all four ROIs. While SDPS produces results that are more closely aligned with the prior distribution, the instability of the algorithm is problematic for the low fidelity (low dose) data simulated in this work. The JSDPS algorithm outperforms both MBMD and SDPS from both visual observations of anatomical structures faithful to the ground truth and according to SSIM and PSNR. Good decomposition results were obtained for both the dual layer and dual kV system, demonstrating generalizability to different spectral systems.

Figure \ref{fig:res_std} shows the bias and standard deviation maps computed from an ensemble of 16 individual output samples. SDPS exhibits large variability and bias around edge features, suggesting substantial image uncertainty. In contrast, JSDPS displays markedly reduced bias and standard deviation, demonstrating the capability to effectively stabilize the sampling process.

\subsubsection{Computational Cost}
Table~\ref{tab:comp} summarizes the computational cost of different material decomposition algorithms for eight output estimates. MBMD is the slowest due to the large number of iterations (10000) required for convergence. SDPS substantially reduces the number of iterations (1000), thereby shortening the computation time. Leveraging the jumpstarted sampling strategy and the gradient approximation, JSDPS requires only 150 iterations, which results in an additional $86.83\%$ and $88.07\%$ reduction in computation time for dual-layer and dual kVp CT, respectively. Furthermore, a $23.16\%$ and $42.54\%$ memory saving was also achieved compared to SDPS for the two spectral systems due to the absence of Jacobian computation.

\begin{table}[h]
\caption{Computational cost of material decomposition algorithms}
\label{tab:comp}
\centering
\begin{tabular}{|c|c|c|c|c|}
\hline
             & \multicolumn{2}{c|}{Dual layer} & \multicolumn{2}{c|}{Dual kVp} \\
\hline
             & Time(s)& Memory(GB) & Time(s) & Memory(GB) \\
\hline
MBMD         & 6406.3 & 11.3 & 2802.5 & 4.8 \\
SDPS         & 728.2  & 17.7 & 478.6  & 13.4 \\
JSDPS        & 95.9   & 13.6 & 57.1   & 7.7 \\
\hline
\end{tabular}
\end{table}

\section{Conclusion and Discussion}
\label{sec:con}
Spectral CT material decomposition is a challenging and ill-conditioned problem, often exhibiting excessive noise and slow convergence. The novel SDPS framework is designed to tackle these issues - combining a sophisticated learned SGM prior and a physical model, to achieve low image noise while enhancing image quality. We additionally adopted jumpstarted sampling and gradient approximation strategies which were found to be effective in stabilizing performance as well as reducing computation time and memory consumption. In evaluations on both dual-layer and dual-kVp systems, JSDPS achieves the highest accuracy, the lowest uncertainty, and the lowest computational costs compared to SDPS and MBMD. This work demonstrated that JSDPS is a promising approach for spectral CT decomposition. 

\section*{Acknowledgments}
This work is supported, in part, by NIH grant R01EB030494.

\bibliography{report}
\bibliographystyle{IEEEref}

\end{document}